\documentclass{sigchi-ext}

\copyrightinfo{Presented at the workshop "Moral Agents for Sustainable Transitions", April 23, 2023, Hamburg, Germany}

\usepackage[english]{babel}

\usepackage[T1]{fontenc}
\usepackage{textcomp}
\usepackage[scaled=.92]{helvet} 
\usepackage{graphicx} 
\usepackage{balance}  
\usepackage{booktabs} 
\usepackage{ccicons}  
\usepackage{ragged2e} 
\usepackage{dirtytalk} 
\usepackage[utf8]{inputenc}  
\usepackage{todonotes} 
\usepackage{float} 
\usepackage{subcaption} 
\usepackage{txfonts}

\usepackage{url}
\makeatletter
\g@addto@macro{\UrlBreaks}{\UrlOrds}
\makeatother

\usepackage{multirow} 
\usepackage{tabularx} 

\usepackage{color} 

\usepackage{enumitem} 

\usepackage[official]{eurosym}
\usepackage{siunitx} 

\usepackage[normalem]{ulem}

\usepackage{microtype}        
\usepackage[all]{hypcap}      
\usepackage{ccicons}          

\usepackage{cite} 

\def\plaintitle{Balancing the Digital and the Physical: Discussing Push and Pull Factors for Digital Well-being} 
  \def\plainauthor{Luca-Maxim Meinhardt, Jan-Henry Belz, Michael Rietzler, Enrico Rukzio}

\def\plainkeywords{digital well-being, interventions, smartphone, nudging, dark patterns, push and pull factors}

\title{\plaintitle}

\numberofauthors{3}
\author{
  \alignauthor{%
    \textbf{Luca-Maxim Meinhardt}\\
    \affaddr{Institute of Media Informatics} \\
    \affaddr{Ulm University, Germany} \\
    \email{luca.meinhardt@uni-ulm.de} } \vfil 
    \alignauthor{%
    \textbf{Jan-Henry Belz}\\    
    \affaddr{Institute of Media Informatics} \\
    \affaddr{Ulm University, Germany} \\
    \email{jan.belz@uni-ulm.de} }\vfil 
    \alignauthor{%
    \textbf{Michael Rietzler}\\    
    \affaddr{Institute of Media Informatics} \\
    \affaddr{Ulm University, Germany} \\
    \email{michael.rietzler@uni-ulm.de} } \vfil
    \alignauthor{%
    \textbf{Enrico Rukzio}\\    
    \affaddr{Institute of Media Informatics} \\
    \affaddr{Ulm University, Germany} \\
    \email{enrico.rukzio@uni-ulm.de} }
    }

\definecolor{linkColor}{RGB}{6,125,233}
\hypersetup{%
  pdftitle={\plaintitle},
  pdfauthor={\plainauthor},
  pdfkeywords={\plainkeywords},
  bookmarksnumbered,
  pdfstartview={FitH},
  colorlinks,
  citecolor=black,
  filecolor=black,
  linkcolor=black,
  urlcolor=linkColor,
  breaklinks=true,
}

\begin{document}


\maketitle

\RaggedRight{} 

\begin{abstract}
This position paper discusses the negative effects of excessive smartphone usage on mental health and well-being. Despite efforts to limit smartphone usage, users become desensitized to reminders and limitations. The paper proposes the use of Push \& Pull Factors to contextualize intervention strategies promoting digital well-being. Further, alternative metrics to measure the effectiveness of intervention strategies will be discussed.
\end{abstract}

\keywords{\plainkeywords}

\ccsdesc[500]{Human-centered computing~HCI theory, concepts and models}
\printccsdesc

\newpage
\section{Introduction}\label{sec:introduction}

Technology integration into daily life has proven to be a double-edged sword. On the one hand, it has greatly facilitated access to information, communication, and entertainment and has thus become a staple of modern existence. On the other hand, its pervasive influence has also brought negative consequences, particularly regarding addiction and its impact on mental health and well-being. The excessive utilization of smartphones, in particular, has been found to have detrimental effects on an individual's psychological state \cite{bashir2017effects, Thomee.2018, Panova.2018}. Moreover, the increasing presence of technology in social settings has resulted in a decline in face-to-face communication, potentially affecting the quality and depth of interpersonal relationships \cite{Przybylski.2013, Lutz.2020, Cesareo.2021, Buttner.2022}. Research on social media has revealed that excessive usage results in symptoms of anxiety, depression \cite{eightyear_depression}, Fear of Missing Out \cite{Brown.2020}, and body dissatisfaction \cite{Fardouly.2016b}. 

Many users are aware of these effects and are willing to change their habits \cite{ko2015nugu} or limit their smartphone usage \cite{hiniker2016mytime} for example by native digital well-being apps for Android and iOS \footnote{ https://www.android.com/digital-wellbeing/ \newline https://support.apple.com/en-us/HT208982}. Despite these efforts, users become desensitized to reminders and limitations, resulting in unconscious and frequent smartphone usage \cite{Almourad.2021}.
This finding is supported by Duke and Montag \cite{Duke.2017}, who found that people tend to underestimate the frequency of their smartphone usage due to the automatic and unconscious nature of this behavior, leading to impulsive actions and distraction. This missing \textit{"feeling of control over actions and their consequences"} \cite[p. 1]{Moore_2017_sense_of_agency} is referred to as the sense of agency.

Research on the effects of phone abstinence has been increasing in response to the above-mentioned effects of excessive smartphone usage \cite{radtke2022digital}. Studies found positive effects of \textit{digital detoxing}, such as reduced stress~\cite{anrijs2018mobiledna}, and depression \cite{Lambert.2022}, as well as increased social connectedness, mental well-being \cite{brown2020fear} and life satisfaction \cite{fioravanti2020taking}. However, there is also research that found negative effects of social media reduction on active users, while passive users did not show effects on interventions Hanley et al. \cite{hanley2019taking}. The restriction of phones could even cause withdrawal symptoms in the first days of reduction \cite{eide2018smartphone}. Therefore, according to Brailovskaia et al. \cite{Brailovskaia.2022}, it is important to find the correct amount of intervention as \textit{"a moderate and controlled reduction of daily time spent on smartphone use could be even more effective and provide more stable effects over a longer period of time."}\cite[p. 11]{Brailovskaia.2022}.

\smallskip
This position paper aims to discuss the effectiveness of intervention strategies promoting digital well-being and to identify gaps in current measurement methodologies. To provide a new perspective on intervention strategies, we propose to bring them into context with \textit{Pull \& Push-Factors}. While these factors have been used in migration theory \cite{push_and_pull}, they have not yet been applied in the context of digital well-being. 

\section{Intervention strategies}
The reduction of excessive smartphone usage can be approached through three strategies, according to Busch et al. \cite{busch2021antecedents}. These strategies, known as \textit{information-enhancing}, \textit{capacity-enhancing}, and \textit{behavior-reinforcement} strategies, are considered as increasing levels of intervention. \textit{Information-enhancing} strategies aim to increase awareness of the usage's negative effects, while \textit{capacity-enhancing} strategies involve providing concrete warnings of destructive usage. \textit{Behavior-reinforcement} strategies address the issue \textit{"through intentional design of usability features that directly constrain users by shutting them completely out of an app or the smartphone after a certain period of time or making smartphone use more difficult"} \cite[p. 14]{busch2021antecedents}.

\subsection{Push \& Pull Factors}
Creating impediments to smartphone usage, such as restricting access to apps or reducing their usability, however, presents a quandary between promoting optimal user experience and addressing the issue of excessive and unhealthy smartphone usage. One solution to the problem could be the concept of \textit{Pleasurable Troublemakers} introduced by Hassenzahl and Laschke \cite{hassenzahl2015pleasurable}. They created little friction in daily routines to unconsciously change people's behavior towards healthy life habits. Hence, applying this mechanism to smartphone intervention showed that smartphone time can be reduced due to discouraging the users \cite{Kim.2019b, Kim.2019, Okeke.2018, Park.2021, ko2015nugu}. Creating friction in the usability or locking them out of an app can be considered as \textit{push factors} as it pushes the user away from the phone. However, according to \cite{Almoallim_2022}, smartphone interventions instead should \textit{"[support] meaningful use rather than limiting meaningless use"}\cite[p. 1]{Almoallim_2022}. Lukoff et al. \cite{Lukoff_2018} supported this quote by stating that meaningful usage should preserve the user's autonomy. Hence, besides creating friction in the smartphone's usability, we should also consider \textit{pull factors}, which pull the users towards meaningful smartphone usage (e.g., mobile learning applications~\cite{Draxler.2019}) to reduce meaningless usage \cite{Terzimehic.2022, Terzimehic.2022b}. Further, physical hardware alternatives that pull users back to the real world, such as proposed by Choi et al. \cite{choi2016lockdoll}, may be even more effective than solely relying on software. This is because the source of distraction is the smartphone itself. Therefore, physical ambient displays also act as a \textit{pull factor} to draw the user's attention back to the physical world instead of the screens~\cite{Khot.2022, Cho.2017}.

\section{Assessment of Interventions}
Both \textit{push} and \textit{pull} factors are important tools in increasing digital well-being and reducing unhealthy screen time. However, several strategies referenced above have been evaluated through user studies, primarily based on reducing screen time. This metric seems to be the predominant metric for the success of an intervention strategy in this research area. However, scholars have raised concerns about the limited representation of users' preferences for digital well-being by this metric \cite{Lukoff_online, Hiniker.2016, Lukoff_2018, Almoallim_2022}. Additionally, \textit{push factors}, which involve intervening with the user's intention, can threaten personal autonomy, leading to reactance and reduced acceptance of the intervention. \textit{"It would, therefore, be undesirable for the creators of technical systems if users would become reactant while using these systems [...]"} \cite[p. 4]{Ehrenbrink.2020}. To address this issue, Ehrenbrink \cite{Ehrenbrink.2020} developed a reactance scale for HCI (RSHCI) to measure users' friction with technical systems.

We propose that the RSHCI can serve as a valuable supplement for evaluating the success of intervention strategies. Effectiveness is an important criterion for interventions, but minimizing the user's reactance is also crucial. For instance, while forcing the phone to turn off after a specific time may be highly effective in reducing screen time, it may also result in high reactance and low acceptance among users. We, therefore, suggest that researchers must strike a balance between the effectiveness of the intervention strategy and users' reactance to create a sustainable increase in digital well-begin.

\section{Outlook and Conclusion}

In this position paper, we aimed to discuss the adverse impacts of excessive smartphone usage on mental health and well-being. Despite the efforts to restrict smartphone usage, users frequently become immune to reminders and limitations. Thus, the paper proposed using Push and Pull Factors to contextualize intervention strategies. It was suggested that in addition to intervention strategies that create friction in the smartphone's usability (push factors), new interventions should also consider factors that attract users to meaningful usage or draw them back to the physical world (pull factors).
Furthermore, the RSHCI, was discussed to measure the success and acceptance of such interventions as a supplement to screen time.

Looking ahead to the near future, Extended Reality (XR) is poised to become ubiquitous, akin to smartphones in our present day. However, besides new opportunities, the pervasiveness of XR devices may introduce new moral challenges \cite{Rixen1, Rixen2}. As these devices have the potential to replace smartphones, there is a risk of encountering similar addiction, distraction, and mental health issues. Hence, it is necessary to expand research on digital well-being and intervention strategies to future technologies to gain a better understanding and address potential challenges.

\balance{} 

\bibliographystyle{SIGCHI-Reference-Format}
\bibliography{sample.bib}


\begin{thebibliography}{00}


\ifx \showCODEN    \undefined \def \showCODEN     #1{\unskip}     \fi
\ifx \showDOI      \undefined \def \showDOI       #1{{\tt DOI:}\penalty0{#1}\ }
  \fi
\ifx \showISBNx    \undefined \def \showISBNx     #1{\unskip}     \fi
\ifx \showISBNxiii \undefined \def \showISBNxiii  #1{\unskip}     \fi
\ifx \showISSN     \undefined \def \showISSN      #1{\unskip}     \fi
\ifx \showLCCN     \undefined \def \showLCCN      #1{\unskip}     \fi
\ifx \shownote     \undefined \def \shownote      #1{#1}          \fi
\ifx \showarticletitle \undefined \def \showarticletitle #1{#1}   \fi
\ifx \showURL      \undefined \def \showURL       #1{#1}          \fi

\bibitem{Almoallim_2022}
{Sultan Almoallim} {and} {Corina Sas}. 2022.
\newblock \showarticletitle{Toward Research-Informed Design Implications for
  Interventions Limiting Smartphone Use: Functionalities Review of Digital
  Well-being Apps}.
\newblock {\em JMIR Form Res\/} {6}, 4 (19 Apr 2022), e31730.
\newblock
\showISSN{2561-326X}
\showDOI{%
\url{http://dx.doi.org/10.2196/31730}}


\bibitem{Almourad.2021}
{Mohamed~Basel Almourad}, {Amen Alrobai}, {Tiffany Skinner}, {Mohammed
  Hussain}, {and} {Raian Ali}. 2021.
\newblock \showarticletitle{Digital wellbeing tools through users lens}.
\newblock {\em Technology in Society\/}  {67} (2021), 101778.
\newblock
\showISSN{0160791X}
\showDOI{%
\url{http://dx.doi.org/10.1016/j.techsoc.2021.101778}}


\bibitem{anrijs2018mobiledna}
{Sarah Anrijs}, {Klaas Bombeke}, {Wouter Durnez}, {Kristin {van Damme}}, {Bart
  Vanhaelewyn}, {Peter Conradie}, {Elena Smets}, {Jan Cornelis}, {Walter de
  Raedt}, {Koen Ponnet}, {and} {Lieven de Marez}. 2018.
\newblock \showarticletitle{MobileDNA: Relating Physiological Stress
  Measurements to Smartphone Usage to Assess the Effect of a Digital Detox}.
\newblock In {\em HCI International 2018 -- Posters' Extended Abstracts},
  {Constantine Stephanidis} (Ed.). Communications in Computer and Information
  Science, Vol. 851. {Springer International Publishing}, Cham, 356--363.
\newblock
\showISBNx{978-3-319-92278-2}
\showDOI{%
\url{http://dx.doi.org/10.1007/978-3-319-92279-9 {\_}48}}


\bibitem{bashir2017effects}
{Hilal Bashir} {and} {Shabir~Ahmad Bhat}. 2017.
\newblock \showarticletitle{Effects of social media on mental health: A
  review}.
\newblock {\em International Journal of Indian Psychology\/} {4}, 3 (2017),
  125--131.
\newblock
\showDOI{%
\url{http://dx.doi.org/10.25215/0403.134}}


\bibitem{Brailovskaia.2022}
{Julia Brailovskaia}, {Jasmin Delveaux}, {Julia John}, {Vanessa Wicker}, {Alina
  Noveski}, {Seokyoung Kim}, {Holger Schillack}, {and} {J{\"u}rgen Margraf}.
  2022.
\newblock \showarticletitle{Finding the {\textquotedbl}sweet
  spot{\textquotedbl} of smartphone use: Reduction or abstinence to increase
  well-being and healthy lifestyle?! An experimental intervention study}.
\newblock {\em Journal of experimental psychology. Applied\/} (2022).
\newblock
\showDOI{%
\url{http://dx.doi.org/10.1037/xap0000430}}


\bibitem{Brown.2020}
{Lorna Brown} {and} {Daria~J. Kuss}. 2020a.
\newblock \showarticletitle{Fear of Missing Out, Mental Wellbeing, and Social
  Connectedness: A Seven-Day Social Media Abstinence Trial}.
\newblock {\em International journal of environmental research and public
  health\/} {17}, 12 (2020).
\newblock
\showDOI{%
\url{http://dx.doi.org/10.3390/ijerph17124566}}


\bibitem{brown2020fear}
{Lorna Brown} {and} {Daria~J. Kuss}. 2020b.
\newblock \showarticletitle{Fear of Missing Out, Mental Wellbeing, and Social
  Connectedness: A Seven-Day Social Media Abstinence Trial}.
\newblock {\em International journal of environmental research and public
  health\/} {17}, 12 (2020).
\newblock
\showDOI{%
\url{http://dx.doi.org/10.3390/ijerph17124566}}


\bibitem{busch2021antecedents}
{Peter~Andr{\'e} Busch} {and} {Stephen McCarthy}. 2021.
\newblock \showarticletitle{Antecedents and consequences of problematic
  smartphone use: A systematic literature review of an emerging research area}.
\newblock {\em Computers in Human Behavior\/}  {114} (2021), 106414.
\newblock
\showISSN{07475632}
\showDOI{%
\url{http://dx.doi.org/10.1016/j. chb.2020.106414}}


\bibitem{Buttner.2022}
{Christiane~M. B{\"u}ttner}, {Andrew~T. Gloster}, {and} {Rainer Greifeneder}.
  2022.
\newblock \showarticletitle{Your phone ruins our lunch: Attitudes, norms, and
  valuing the interaction predict phone use and phubbing in dyadic social
  interactions}.
\newblock {\em Mobile Media {\&} Communication\/} {10}, 3 (2022), 387--405.
\newblock
\showISSN{2050-1579}
\showDOI{%
\url{http://dx.doi.org/10.1177/20501579211059914}}


\bibitem{Cesareo.2021}
{Massimo Cesareo}, {Marco Tagliabue}, {Annalisa Oppo}, {and} {Paolo Moderato}.
  2021.
\newblock \showarticletitle{The ubiquity of social reinforcement: A nudging
  exploratory study to reduce the overuse of smartphones in social contexts}.
\newblock {\em Cogent Psychology\/} {8}, 1 (2021).
\newblock
\showDOI{%
\url{http://dx.doi.org/10.1080/23311908.2021.1880304}}


\bibitem{Cho.2017}
{Minjoo Cho} {and} {Daniel Saakes}. 2017.
\newblock \showarticletitle{Calm Automaton}. In {\em Proceedings of the 2017
  CHI Conference Extended Abstracts on Human Factors in Computing Systems},
  {Gloria Mark}, {Susan Fussell}, {Cliff Lampe}, {m.c. schraefel}, {Juan~Pablo
  Hourcade}, {Caroline Appert}, {and} {Daniel Wigdor} (Eds.). ACM, New York,
  NY, USA, 393--396.
\newblock
\showISBNx{9781450346566}
\showDOI{%
\url{http://dx.doi.org/10.1145/3027063.3052968}}


\bibitem{choi2016lockdoll}
{Seungwoo Choi}, {Hayeon Jeong}, {Minsam Ko}, {and} {Uichin Lee}. 2016.
\newblock \showarticletitle{LockDoll}. In {\em Proceedings of the 2016 CHI
  Conference Extended Abstracts on Human Factors in Computing Systems}, {Jofish
  Kaye}, {Allison Druin}, {Cliff Lampe}, {Dan Morris}, {and} {Juan~Pablo
  Hourcade} (Eds.). ACM, New York, NY, USA, 1165--1172.
\newblock
\showISBNx{9781450340823}
\showDOI{%
\url{http://dx.doi.org/10.1145/2851581.2892445}}


\bibitem{eightyear_depression}
{Sarah~M. Coyne}, {Adam~A. Rogers}, {Jessica~D. Zurcher}, {Laura Stockdale},
  {and} {McCall Booth}. 2020.
\newblock \showarticletitle{Does time spent using social media impact mental
  health?: An eight year longitudinal study}.
\newblock {\em Computers in Human Behavior\/}  {104} (2020), 106160.
\newblock
\showISSN{0747-5632}
\showDOI{%
\url{http://dx.doi.org/https://doi.org/10.1016/j .chb.2019.106160}}


\bibitem{Draxler.2019}
{Fiona Draxler}, {Christina Schneegass}, {and} {Evangelos Niforatos}. 2019.
\newblock \showarticletitle{Designing for Task Resumption Support in Mobile
  Learning}. In {\em Proceedings of the 21st International Conference on
  Human-Computer Interaction with Mobile Devices and Services}. ACM, New York,
  NY, USA, 1--6.
\newblock
\showISBNx{9781450368254}
\showDOI{%
\url{http://dx.doi.org/10.1145/3338286.3344394}}


\bibitem{Duke.2017}
{{\'E}ilish Duke} {and} {Christian Montag}. 2017.
\newblock \showarticletitle{Smartphone addiction, daily interruptions and
  self-reported productivity}.
\newblock {\em Addictive behaviors reports\/}  {6} (2017), 90--95.
\newblock
\showDOI{%
\url{http://dx.doi.org/10.1016/j. abrep.2017.07.002}}


\bibitem{Ehrenbrink.2020}
{Patrick Ehrenbrink}. 2020.
\newblock \showarticletitle{Reactance Scale for Human--Computer Interaction}.
\newblock In {\em The Role of Psychological Reactance in Human--Computer
  Interaction}, {Patrick Ehrenbrink} (Ed.). {Springer International
  Publishing}, Cham, 71--81.
\newblock
\showISBNx{978-3-030-30309-9}
\showDOI{%
\url{http://dx.doi.org/10.1007/978-3-030-30310-5 {\_}8}}


\bibitem{eide2018smartphone}
{Tine~A. Eide}, {Sarah~H. Aarestad}, {Cecilie~S. Andreassen}, {Robert~M.
  Bilder}, {and} {St{\aa}le Pallesen}. 2018.
\newblock \showarticletitle{Smartphone Restriction and Its Effect on Subjective
  Withdrawal Related Scores}.
\newblock {\em Frontiers in psychology\/}  {9} (2018), 1444.
\newblock
\showISSN{1664-1078}
\showDOI{%
\url{http://dx.doi.org/10.3389/fpsyg.2018.01444}}


\bibitem{Fardouly.2016b}
{Jasmine Fardouly} {and} {Lenny~R. Vartanian}. 2016.
\newblock \showarticletitle{Social Media and Body Image Concerns: Current
  Research and Future Directions}.
\newblock {\em Current Opinion in Psychology\/}  {9} (2016), 1--5.
\newblock
\showISSN{2352250X}
\showDOI{%
\url{http://dx.doi.org/10.1016/j. copsyc.2015.09.005}}


\bibitem{fioravanti2020taking}
{Giulia Fioravanti}, {Alfonso Prostamo}, {and} {Silvia Casale}. 2020.
\newblock \showarticletitle{Taking a Short Break from Instagram: The Effects on
  Subjective Well-Being}.
\newblock {\em Cyberpsychology, behavior and social networking\/} {23}, 2
  (2020), 107--112.
\newblock
\showDOI{%
\url{http://dx.doi.org/10.1089/cyber.2019.0400}}


\bibitem{hanley2019taking}
{Sarah~M. Hanley}, {Susan~E. Watt}, {and} {William Coventry}. 2019.
\newblock \showarticletitle{Taking a break: The effect of taking a vacation
  from Facebook and Instagram on subjective well-being}.
\newblock {\em PloS one\/} {14}, 6 (2019), e0217743.
\newblock
\showDOI{%
\url{http://dx.doi.org/10.1371/journal.pone.0217743}}


\bibitem{hassenzahl2015pleasurable}
{Marc Hassenzahl} {and} {Matthias Laschke}. 2015.
\newblock \showarticletitle{Pleasurable Troublemakers}.
\newblock {\em The Gameful World: Approaches, Issues, Applications\/} (2015),
  167.
\newblock


\bibitem{Hiniker.2016}
{Alexis Hiniker}, {Sungsoo Hong}, {Tadayoshi Kohno}, {and} {Julie~A. Kientz}.
  2016a.
\newblock \showarticletitle{MyTime}. In {\em Proceedings of the 2016 CHI
  Conference on Human Factors in Computing Systems}, {Jofish Kaye}, {Allison
  Druin}, {Cliff Lampe}, {Dan Morris}, {and} {Juan~Pablo Hourcade} (Eds.). ACM,
  New York, NY, USA, 4746--4757.
\newblock
\showISBNx{9781450333627}
\showDOI{%
\url{http://dx.doi.org/10.1145/2858036.2858403}}


\bibitem{hiniker2016mytime}
{Alexis Hiniker}, {Sungsoo Hong}, {Tadayoshi Kohno}, {and} {Julie~A. Kientz}.
  2016b.
\newblock \showarticletitle{MyTime: Designing and Evaluating an Intervention
  for Smartphone Non-Use}. In {\em Proceedings of the 2016 CHI Conference on
  Human Factors in Computing Systems} {\em (CHI '16)}. {Association for
  Computing Machinery}, New York, NY, USA, 4746--4757.
\newblock
\showISBNx{9781450333627}
\showDOI{%
\url{http://dx.doi.org/10.1145/2858036.2858403}}


\bibitem{Khot.2022}
{Rohit~Ashok Khot}, {Jung-Ying Yi}, {and} {Deepti Aggarwal}. 2022.
\newblock \showarticletitle{Designing for Microbreaks: Unpacking the Design
  Journey of Zenscape}. In {\em Sixteenth International Conference on Tangible,
  Embedded, and Embodied Interaction}. ACM, New York, NY, USA, 1--16.
\newblock
\showISBNx{9781450391474}
\showDOI{%
\url{http://dx.doi.org/10.1145/3490149.3502256}}


\bibitem{Kim.2019}
{Jaejeung Kim}, {Hayoung Jung}, {Minsam Ko}, {and} {Uichin Lee}. 2019a.
\newblock \showarticletitle{GoalKeeper}.
\newblock {\em Proceedings of the ACM on Interactive, Mobile, Wearable and
  Ubiquitous Technologies\/} {3}, 1 (2019), 1--29.
\newblock
\showDOI{%
\url{http://dx.doi.org/10.1145/3314403}}


\bibitem{Kim.2019b}
{Jaejeung Kim}, {Joonyoung Park}, {Hyunsoo Lee}, {Minsam Ko}, {and} {Uichin
  Lee}. 2019b.
\newblock \showarticletitle{LocknType}. In {\em Proceedings of the 2019 CHI
  Conference on Human Factors in Computing Systems}, {Stephen Brewster},
  {Geraldine Fitzpatrick}, {Anna Cox}, {and} {Vassilis Kostakos} (Eds.). ACM,
  New York, NY, USA, 1--12.
\newblock
\showISBNx{9781450359702}
\showDOI{%
\url{http://dx.doi.org/10.1145/3290605.3300927}}


\bibitem{ko2015nugu}
{Minsam Ko}, {Subin Yang}, {Joonwon Lee}, {Christian Heizmann}, {Jinyoung
  Jeong}, {Uichin Lee}, {Daehee Shin}, {Koji Yatani}, {Junehwa Song}, {and}
  {Kyong-Mee Chung}. 2015.
\newblock \showarticletitle{NUGU}. In {\em Proceedings of the 18th ACM
  Conference on Computer Supported Cooperative Work {\&} Social Computing},
  {Dan Cosley}, {Andrea Forte}, {Luigina Ciolfi}, {and} {David McDonald}
  (Eds.). ACM, New York, NY, USA, 1235--1245.
\newblock
\showISBNx{9781450329224}
\showDOI{%
\url{http://dx.doi.org/10.1145/2675133.2675244}}


\bibitem{Lambert.2022}
{Jeffrey Lambert}, {George Barnstable}, {Eleanor Minter}, {Jemima Cooper},
  {and} {Desmond McEwan}. 2022.
\newblock \showarticletitle{Taking a One-Week Break from Social Media Improves
  Well-Being, Depression, and Anxiety: A Randomized Controlled Trial}.
\newblock {\em Cyberpsychology, behavior and social networking\/} {25}, 5
  (2022), 287--293.
\newblock
\showDOI{%
\url{http://dx.doi.org/10.1089/cyber.2021.0324}}


\bibitem{push_and_pull}
{Everett~S. Lee}. 1966.
\newblock \showarticletitle{{A theory of migration}}.
\newblock {\em Demography\/} {3}, 1 (03 1966), 47--57.
\newblock
\showISSN{0070-3370}
\showDOI{%
\url{http://dx.doi.org/10.2307/2060063}}


\bibitem{Lukoff_online}
{Kai Lukoff}. 2019.
\newblock Digital wellbeing is way more than just reducing screen time.
\newblock   (jul 2019).
\newblock
\showURL{%
Retrieved Feb 16, 2023 from
  \url{https://uxdesign.cc/digital-wellbeing-more-than-just-reducing-screen-time-46223db9f057}}


\bibitem{Lukoff_2018}
{Kai Lukoff}, {Cissy Yu}, {Julie Kientz}, {and} {Alexis Hiniker}. 2018.
\newblock \showarticletitle{What Makes Smartphone Use Meaningful or
  Meaningless?}
\newblock {\em Proc. ACM Interact. Mob. Wearable Ubiquitous Technol.\/} {2}, 1,
  Article 22 (mar 2018), 26 pages.
\newblock
\showDOI{%
\url{http://dx.doi.org/10.1145/3191754}}


\bibitem{Lutz.2020}
{Sarah Lutz} {and} {Karin Knop}. 2020.
\newblock \showarticletitle{Put down your smartphone -- unless you integrate it
  into the conversation! An experimental investigation of using smartphones
  during face to face communication}.
\newblock {\em Studies in Communication and Media\/} {9}, 4 (2020), 516--539.
\newblock
\showISSN{2192-4007}
\showDOI{%
\url{http://dx.doi.org/10.5771/2192-4007-2020-4-516}}


\bibitem{Moore_2017_sense_of_agency}
{James~W. Moore}. 2016.
\newblock \showarticletitle{What Is the Sense of Agency and Why Does it
  Matter?}
\newblock {\em Frontiers in Psychology\/}  {7} (2016).
\newblock
\showISSN{1664-1078}
\showDOI{%
\url{http://dx.doi.org/10.3389/fpsyg.2016.01272}}


\bibitem{Okeke.2018}
{Fabian Okeke}, {Michael Sobolev}, {Nicola Dell}, {and} {Deborah Estrin}. 2018.
\newblock \showarticletitle{Good vibrations}. In {\em Proceedings of the 20th
  International Conference on Human-Computer Interaction with Mobile Devices
  and Services}, {Lynne Bailie} {and} {Nuria Oliver} (Eds.). ACM, New York, NY,
  USA, 1--12.
\newblock
\showISBNx{9781450358989}
\showDOI{%
\url{http://dx.doi.org/10.1145/3229434.3229463}}


\bibitem{Panova.2018}
{Tayana Panova} {and} {Xavier Carbonell}. 2018.
\newblock \showarticletitle{Is smartphone addiction really an addiction?}
\newblock {\em Journal of behavioral addictions\/} {7}, 2 (2018), 252--259.
\newblock
\showDOI{%
\url{http://dx.doi.org/10.1556/2006.7.2018.49}}


\bibitem{Park.2021}
{Joonyoung Park}, {Hyunsoo Lee}, {Sangkeun Park}, {Kyong-Mee Chung}, {and}
  {Uichin Lee}. 2021.
\newblock \showarticletitle{GoldenTime: Exploring System-Driven Timeboxing and
  Micro-Financial Incentives for Self-Regulated Phone Use}. In {\em Proceedings
  of the 2021 CHI Conference on Human Factors in Computing Systems}, {Yoshifumi
  Kitamura}, {Aaron Quigley}, {Katherine Isbister}, {Takeo Igarashi}, {Pernille
  Bj{\o}rn}, {and} {Steven Drucker} (Eds.). ACM, New York, NY, USA, 1--17.
\newblock
\showISBNx{9781450380966}
\showDOI{%
\url{http://dx.doi.org/10.1145/3411764.3445489}}


\bibitem{Przybylski.2013}
{Andrew~K. Przybylski} {and} {Netta Weinstein}. 2013.
\newblock \showarticletitle{Can you connect with me now? How the presence of
  mobile communication technology influences face-to-face conversation
  quality}.
\newblock {\em Journal of Social and Personal Relationships\/} {30}, 3 (2013),
  237--246.
\newblock
\showISSN{0265-4075}


\bibitem{radtke2022digital}
{Theda Radtke}, {Theresa Apel}, {Konstantin Schenkel}, {Jan Keller}, {and}
  {Eike von Lindern}. 2022.
\newblock \showarticletitle{Digital detox: An effective solution in the
  smartphone era? A systematic literature review}.
\newblock {\em Mobile Media {\&} Communication\/} {10}, 2 (2022), 190--215.
\newblock
\showISSN{2050-1579}
\showDOI{%
\url{http://dx.doi.org/10.1177/20501579211028647}}


\bibitem{Rixen2}
{Jan~Ole Rixen}, {Mark Colley}, {Ali Askari}, {Jan Gugenheimer}, {and} {Enrico
  Rukzio}. 2022.
\newblock \showarticletitle{Consent in the Age of AR: Investigating The Comfort
  With Displaying Personal Information in Augmented Reality}. In {\em
  Proceedings of the 2022 CHI Conference on Human Factors in Computing Systems}
  {\em (CHI '22)}. Association for Computing Machinery, New York, NY, USA,
  Article 295, 14 pages.
\newblock
\showISBNx{9781450391573}
\showDOI{%
\url{http://dx.doi.org/10.1145/3491102.3502140}}


\bibitem{Rixen1}
{Jan~Ole Rixen}, {Teresa Hirzle}, {Mark Colley}, {Yannick Etzel}, {Enrico
  Rukzio}, {and} {Jan Gugenheimer}. 2021.
\newblock \showarticletitle{Exploring Augmented Visual Alterations in
  Interpersonal Communication}. In {\em Proceedings of the 2021 CHI Conference
  on Human Factors in Computing Systems} {\em (CHI '21)}. Association for
  Computing Machinery, New York, NY, USA, Article 730, 11 pages.
\newblock
\showISBNx{9781450380966}
\showDOI{%
\url{http://dx.doi.org/10.1145/3411764.3445597}}


\bibitem{Terzimehic.2022b}
{Na{\dj}a Terzimehi{\'c}} {and} {Sarah Aragon-Hahner}. 2022.
\newblock \showarticletitle{I Wish I Had: Desired Real-World Activities Instead
  of Regretful Smartphone Use}. In {\em Proceedings of the 21st International
  Conference on Mobile and Ubiquitous Multimedia}, {Tanja D{\"o}ring}, {Susanne
  Boll}, {Ashley Colley}, {Augusto Esteves}, {and} {Jo{\~a}o Guerreiro} (Eds.).
  ACM, New York, NY, USA, 47--52.
\newblock
\showISBNx{9781450398206}
\showDOI{%
\url{http://dx.doi.org/10.1145/3568444.3568465}}


\bibitem{Thomee.2018}
{Sara Thom{\'e}e}. 2018.
\newblock \showarticletitle{Mobile Phone Use and Mental Health. A Review of the
  Research That Takes a Psychological Perspective on Exposure}.
\newblock {\em International journal of environmental research and public
  health\/} {15}, 12 (2018).
\newblock
\showDOI{%
\url{http://dx.doi.org/10.3390/ijerph15122692}}


\end{thebibliography}

\end{document}